\ifx\pdfoutput\undefined        
  \documentclass{an}
  \usepackage{url}
\else                           
  \documentclass[pdftex]{an}
  \usepackage[bookmarks=false]{hyperref}
\fi

\usepackage{times}
\usepackage{graphicx}
\usepackage{fancyhdr}
\usepackage{url}
\usepackage{bm}
\newcommand{\EQ}{\begin{equation}}
\newcommand{\EN}{\end{equation}}
\newcommand{\EQA}{\begin{eqnarray}}
\newcommand{\ENA}{\end{eqnarray}}

\newcommand{\Eq}[1]{Eq.~(\ref{#1})}

\newcommand{\Fig}[1]{Fig.~\ref{#1}}
\newcommand{\FFig}[1]{Figure~\ref{#1}}
\newcommand{\Tab}[1]{Table~\ref{#1}}

\newcommand{\bra}[1]{\langle #1\rangle}

\newcommand{\meanBB}{\overline{\vec{B}}}
\newcommand{\meanJJ}{\overline{\vec{J}}}

\newcommand{\meanWW}{\overline{\vec{W}}}

\newcommand{\bp}{\mbox{\boldmath $p$} {}}
\newcommand{\qq}{\mbox{\boldmath $q$} {}}

\newcommand{\uu}{{\vec{u}}}
\newcommand{\BB}{{\vec{B}}}

\newcommand{\jj}{{\vec{j}}}

\newcommand{\bb}{{\vec{b}}}

\newcommand{\kk}{\mbox{\boldmath $k$} {}}

\newcommand{\nab}{\mbox{\boldmath $\nabla$} {}}

\newcommand{\dd}{{\rm d} {}}

\def\half{{\textstyle{1\over2}}}

\newcommand{\G}{\,{\rm G}}

\newcommand{\uG}{\,\mu{\rm G}}

\newcommand{\g}{\,{\rm g}}
\newcommand{\s}{\,{\rm s}}

\newcommand{\cm}{\,{\rm cm}}

\newcommand{\km}{\,{\rm km}}

\newcommand{\kpc}{\,{\rm kpc}}
\newcommand{\yr}{\,{\rm yr}}
\newcommand{\Myr}{\,{\rm Myr}}
\newcommand{\Gyr}{\,{\rm Gyr}}
\newcommand{\erg}{\,{\rm erg}}

\newcommand{\AU}{\,{\rm AU}}

\newcommand{\yjgr}[3]{: #1, {JGR} {#2}, #3}

\newcommand{\yapj}[3]{: #1, {ApJ} {#2}, #3}

\newcommand{\yan}[3]{: #1, {AN} {#2}, #3}

\newcommand{\ymhd}[3]{: #1, {Magnetohydrodynamics} {#2}, #3}
\newcommand{\yqjras}[3]{: #1, {QJRAS} {#2}, #3}
\newcommand{\yrmp}[3]{: #1, {RvMP} {#2}, #3}
\newcommand{\yana}[3]{: #1, {A\&A} {#2}, #3}

\newcommand{\yaraa}[3]{: #1, {ARA\&A} {#2}, #3}

\newcommand{\ygafd}[3]{: #1, {GApFD} {#2}, #3}
\newcommand{\yjfm}[3]{: #1, {JFM} {#2}, #3}
\newcommand{\yplb}[3]{: #1, {PhLB} {#2}, #3}

\newcommand{\ypp}[3]{: #1, {PhPl} {#2}, #3}

\newcommand{\yrpp}[3]{: #1, {RPPh} {#2}, #3}
\newcommand{\yprl}[3]{: #1, {PhRvL} {#2}, #3}
\newcommand{\yprd}[3]{: #1, {PhRvD} {#2}, #3}
\newcommand{\ypre}[3]{: #1, {PhRvE} {#2}, #3}
\newcommand{\yprt}[3]{: #1, {PhR} {#2}, #3}

\newcommand{\yptrs}[3]{: #1, {RSPTA} {#2}, #3}
\newcommand{\ymn}[3]{: #1, {MNRAS} {#2}, #3}

\newcommand{\ysci}[3]{: #1, {Sci} {#2}, #3}
\newcommand{\ysph}[3]{: #1, {Solar Phys.} {#2}, #3}
\newcommand{\ybasi}[3]{: #1, {BASI} {#2}, #3}

\newcommand{\yjour}[4]{: #1, {#2} {#3}, #4}

\newcommand{\yproc}[5]{: #1, in #4 (eds.), {\it #3}, #5, p.~#2}

\newcommand{\sprl}[1]{: #1, {PRL}, submitted}

\newcommand{\pmn}[1]{: #1, {MNRAS}, in press}

\newcommand{\pana}[1]{: #1, {A\&A}, in press}

\title{Magnetic helicity in primordial and dynamo scenarios of galaxies}
\author{Axel Brandenburg}
\institute{
NORDITA, Blegdamsvej 17, DK-2100 Copenhagen \O, Denmark\\
}

\date{Received 10 December 2005; accepted 23 January 2006;
published online...}
\begin{document}

\abstract{
Some common properties of helical magnetic fields in decaying and
driven turbulence are discussed.
These include mainly the inverse cascade that produces fields on progressively
larger scales.
Magnetic helicity also restricts the evolution of the large-scale field:
the field decays less
rapidly than a non-helical field, but it also saturates more slowly, i.e.\
on a resistive time scale if there are no magnetic helicity fluxes.
The former effect is utilized in primordial field scenarios, while
the latter is important for successfully explaining astrophysical dynamos
that saturate faster than resistively.
Dynamo action is argued to be important not only in the galactic dynamo,
but also in accretion discs in active galactic nuclei and around
protostars, both of which contribute to producing a strong enough
seed magnetic field.
Although primordial magnetic fields may be too weak to compete with these
astrophysical mechanisms, such fields could perhaps still be important
in producing polarization effects in the cosmic background radiation.
}

\maketitle

\section{Introduction}

Magnetic helicity plays a fundamental role both in primordial
theories of galactic magnetism as well as in dynamo theories
amplifying and sustaining contemporary galactic fields.
Both issues have been reviewed in recent years (Grasso \& Rubinstein 2001;
Widrow 2002; Giovannini 2004; Brandenburg \& Subramanian 2005a).
We will therefore only try to collect the main points relevant to the
issues concerning magnetic helicity in galactic and protogalactic magnetism.

The main reason magnetic helicity is at all of concern to us is that
even in the resistive case the rate of magnetic helicity dissipation
asymptotes to zero as the magnetic Reynolds number goes to infinity.
This is not the case with magnetic energy dissipation, which remains
always important, and does not decrease with increasing magnetic
Reynolds number (Galsgaard \& Nordlund 1996).
Therefore the magnetic helicity is nearly conserved at all times.
This has serious consequences for the evolution of
magnetohydrodynamic (MHD) turbulence, as has been demonstrated by a number
of recent simulations when the resolution has been large enough
(Brandenburg 2001a; Mininni et al.\ 2005).

At a more descriptive level, magnetic helicity characterizes the
degree of field line linkage.
As the magnetic field relaxes, its energy decreases, but the linkage
stays, at least as much as possible.
The field's inability to undo its knots implies also that the field
cannot decay freely.
This slows down the decay, which is important if a primordial field
is to be of any significance at the time of recombination.
In the driven case, on the other hand, magnetic helicity is better
pictured in terms of writhe and twist
(e.g.\ Longcope \& Klapper 1997; D\'emoulin et al.\ 2002).
Writhe refers to the tilt of a flux tube, and we use both
expressions synonymously.
A cyclonic event tilts individual flux tubes, but as it does so, a
corresponding amount of internal twist is necessarily introduced in the
tube (Blackman \& Brandenburg 2003).
This is what saturates the dynamo, and this can be a very powerful
effect if the small-scale internal twist cannot escape.
In this review we discuss both decaying and driven turbulence.
The former is relevant for prolonging the decay of a primordial
field, while the latter is relevant for understanding how the
galactic dynamo saturates and how to enable it to do so faster.

\section{Magnetic helicity in the primordial scenario}

Theories of the electroweak phase transition, about $10^{-10}\s$
or less after the big bang, allow for the possibility of generating a
magnetic field of up to $10^{24}\G$ (see Grasso \& Rubinstein 2001).
[In practice the field will be weaker; Brandenburg et al.\ (1996a)
discussed a field of $10^{18}\G$ at the time of the electroweak phase
transition which would have decayed to $10^{-11}\G$ at the present time.]
The scale of this field would be less than or comparable to
the horizon scale which was only about $3\cm$ or less.
With the cosmological expansion this field would have a scale of about
$1\AU$, which is still small compared with the scale of galaxies
(Hindmarsh \& Everett 1998).
This led to the idea that the inverse cascade of magnetic helicity
might have played a role in increasing the scale of the turbulent
magnetic field (Brandenburg et al.\ 1996a; Field \& Carroll 2000);
see also Brandenburg (2001b) for a summarizing view.
Therefore we address in this section how a helical magnetic field decays.
The only source of turbulence is assumed to be the initial magnetic field
itself, which drives a flow through the Lorentz force.

\subsection{Scaling of energy spectrum during inverse transfer}

There are indeed certain possibilities for producing primordial magnetic
fields that may have had significant amounts of magnetic helicity
(Joyce \& Shaposhnikov 1997; Cornwall 1997; Vachaspati 2001;
Semikoz \& Sokoloff 2004, 2005).
Letting the field inverse cascade has also the advantageous side effect
that the resulting large-scale fields can more easily overcome Silk
damping during the period of recombination (Brandenburg et al.\ 1997).
This damping was previously thought to be a serious threat to
primordial theories that generated magnetic field during early
Universe phase transitions, but calculations showed that the
Alfv\'en mode can survive for scales smaller than the Silk scale
(Subramanian \& Barrow 1998a; Jedamzik et al.\ 1998).

\begin{figure}[t!]\begin{center}
\includegraphics[width=\columnwidth]{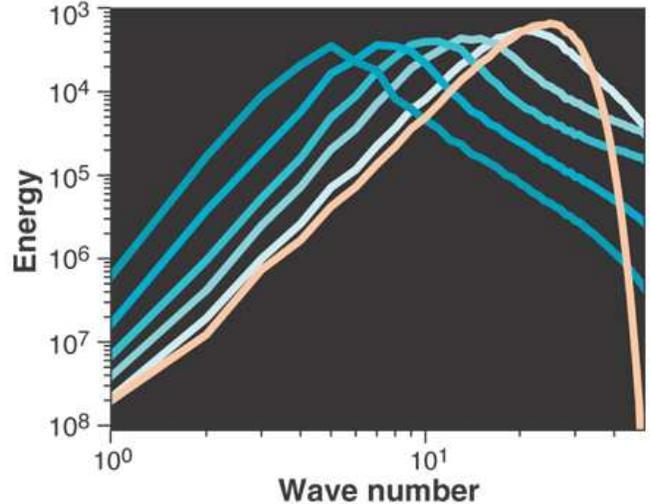}
\end{center}\caption[]{
Magnetic energy spectra at different times
(increasing roughly by a factor of 2).
The curve with the right-most location of the peak corresponds to
the initial time, while the other lines refer to later times (increasing
from right to left). Note the
propagation of spectral energy to successively smaller wavenumbers $k$,
i.e.\ to successively larger scales. [Adapted from Christensson et al.\ (2001)
and Brandenburg (2001b)]
}\label{InvCasc}\end{figure}

More recently simulations have directly been able to demonstrate
how the inverse cascade works.
This can be seen from magnetic power spectra at different times after
initializing the simulation with a random helical magnetic field.
\FFig{InvCasc} demonstrates quite clearly that in decaying turbulence
an inverse cascade means not just that the dominant scale increases,
because any diffusion that is more efficient on smaller scales than on
larger scales must increase the relative dominance of large-scale fields
over small-scale fields.
Instead, inverse cascade means actually a real increase of the field
strength at large scales, i.e.\ the spectral energy $E(k,t)$ increases
with $t$ for $k<k_{\rm peak}$.
Here, $k_{\rm peak}$ is the wavenumber where the magnetic power spectrum
peaks; this value is decreasing with time.

The work of Christensson et al.\ (2001) has also shown that the spectrum
stays approximately shape-invariant; see also Banerjee \& Jedamzik (2004).
Indeed, the time-dependent spectrum can be written as
\EQ
E_{\rm M}(k,t)=\xi(t)^{-q}g_{\rm M}(k\xi),
\label{MagneticScaling}
\EN
where $\xi(t)$ is the characteristic length scale of the magnetic field
and $g_{\rm M}(k\xi)$ is the magnetic scaling function
[see Christensson et al.\ (2001) for a plot of $g_{\rm M}(k\xi)$].
The magnetic energy can then be written as
\EQ
E_{\rm M}(t)=\int_0^{k_{\max}}E_{\rm M}(k,t)\;\dd k\propto\xi^{-(q+1)}.
\EN
Here we have assumed that the spectrum has an upper cutoff at $k_{\max}$,
which is comparable to and probably slightly larger than $k_{\rm peak}$.
Furthermore, if the characteristic length scale changes with time
in a power law fashion, $\xi\propto t^r$, we find a decay law of
magnetic energy like $E_{\rm M}(t)\propto t^n$ with $n=r(q+1)$.
This allows us now to calculate how the spectral energy at large scales
(small values of $k$) depends on time.
For small values of $k$ we now assume that \Eq{MagneticScaling} can be
written in power law form as
\EQ
E_{\rm M}(k,t)=k^pt^\sigma\quad\mbox{(for $k\ll k_{\max}$)}.
\EN
We are interested in the exponent $\sigma$ that tells us how the
spectral magnetic energy grows in time.
Using \Eq{MagneticScaling}, assuming that $g_{\rm M}(k\xi)=(k\xi)^p$,
we have
\EQ
E_{\rm M}(k,t)=\xi(t)^{-q}(k\xi)^p=k^p\xi^{p-q}=k^pt^{r(p-q)}.
\EN
Expressing $q$ in terms of $n$, we find
\EQ
\sigma=r(p+1)-n.
\label{SigmaDep}
\EN
In \Tab{Tsummary} we give the results for different values of $p$, $n$,
and $r$.
The first entry in this table ($p=4$, $n=1/2$, and $r=1/2$) is basically
the case considered by Christensson et al.\ (2005), except that they
also found an additive correction to $n$ (see below), which directly
affects $\sigma$.

\begin{table}[htb]\caption{
Values of $\sigma$ for different combinations of
$p$, $n$, and $r$, as given by \Eq{SigmaDep}.
The first row applies to helical turbulence in the limit
of large magnetic Reynolds numbers with $p=4$.
}\vspace{12pt}\centerline{\begin{tabular}{lccccccc}
$p$ & $n$ & $r$ & $\sigma$ \\
\hline
 4  & 1/2 & 1/2 &  2  \\
 2  & 1/2 & 1/2 &  1  \\
 0  & 1/2 & 1/2 &  0  \\
 4  &  1  & 1/2 & 3/2 \\
 2  &  1  & 1/2 & 1/2 \\
$p$ & $n$ &  0  & $-n$ \\
 4  &  1  &  1  &  4  \\
\label{Tsummary}\end{tabular}}\end{table}

Looking at \Tab{Tsummary} and also at \Eq{SigmaDep}, it is clear that
a steeper spectrum (larger $p$) and a faster increase of the length
scale (larger $r$) yield a faster rise of the spectral power at low
$k<k_{\max}$, while the overall decay exponent, $-n$, directly adds
to $\sigma$.
For the case considered by Christensson et al.\ (2001), where $p=4$,
$n\approx0.7$, $r\approx0.5$, one finds $\sigma\approx1.8$, which is
compatible with the rise of spectral energy (for $k<k_{\rm peak}$)
seen in \Fig{InvCasc}.

\subsection{Simple argument for inverse transfer}

At this point it may be useful to provide a simple argument
[due to Frisch et al.\ (1975)] as to why the interaction of helical
magnetic fields leads preferentially to large-scale magnetic fields.
We reproduce here the argument as presented in the review by
Brandenburg \& Subramanian (2005a).
In this argument one assumes that two waves with wavevectors $\bp$
and $\qq$ interact with each other to produce a wave of wavevector $\kk$.
Both waves are assumed to be fully helical with the same sign of helicity.
Assuming that the total energy $E$ (which is the sum of magnetic and
kinetic energies) is conserved together with magnetic helicity, we have
\EQ
E_p + E_q = E_k ,
\label{frisch_energy}
\EN
\EQ
|H_p| + |H_q| = |H_k|.
\label{helconspqk}
\EN
(Since in this system the flow is driven by the magnetic field, we can ignore
the kinetic energy compared with the magnetic energy, so for all practical
purposes we can think of $E$ being equivalent to $E_{\rm M}$.)
Since both waves are fully helical, we have
\EQ
2E_p=p|H_p|\quad\mbox{and}\quad
2E_q=q|H_q|,
\EN
and so \Eq{frisch_energy} yields
\EQ
p |H_p| + q|H_q| = 2E_k \ge k|H_k| ,
\label{helmod}
\EN
where the last inequality is also known as the realizability condition
that is here applied to the target wavevector $\kk$ after the interaction.
Using \Eq{helconspqk} in \Eq{helmod} we have
\EQ
p|H_p|+q|H_q|\ge k(|H_p|+|H_q|).
\EN
In other words, the target wavevector $\kk$ after the interaction of
wavevectors $\bp$ and $\qq$ satisfies
\EQ
k\le{p|H_p|+q|H_q|\over|H_p|+|H_q|}.
\label{helconspqk2}
\EN
The expression on the right hand side of \Eq{helconspqk2} is a weighted
mean of $p$ and $q$ and thus satisfies
\EQ
\min(p,q)\le{p|H_p|+q|H_q|\over|H_p|+|H_q|}\le\max(p,q),
\EN
and therefore
\EQ
k\le\max(p,q).
\EN
In the special case where $p=q$, we have $k\le p=q$, so the target
wavenumber after interaction is always less than or equal to the initial
wavenumbers. In other words, wave interactions tend to transfer some magnetic
energy to smaller wavenumbers, i.e.\ to larger scale. This corresponds
to an inverse cascade. The realizability condition, $\half k|H_k|\le E_k$,
was the most important ingredient in this argument.
An important assumption that we made in the beginning was that the initial
field be fully helical; see Maron \& Blackman (2002) and
Brandenburg et al.\ (2002) for simulations of driven turbulence with
fractional helicity.

\subsection{Decay law}

The magnetic energy decay is often seen to follow power law behavior,
i.e.\ $E(t)\sim t^{-n}$.
For nonhelical turbulence, $n$ is typically larger than unity
[e.g.\ $n=1.28$ in the work of Mac Low et al.\ (1998),
or $n=1.2$ in the argument discussed by Subramanian et al.\ (2005)].
On the other hand, for helical turbulence the decay is more
shallow; for example Biskamp \& M\"uller (1999) find typical values between
0.5 and 0.7.
They explain their scaling with the following argument.
They assume that the magnetic helicity $H$ is perfectly conserved,
so $H(t)=\mbox{const}$, and so the typical length scale $L(t)$ depends
only on the total energy, $E(t)$, via $L\sim H/E\sim E^{-1}$.
Assuming furthermore that the rate of energy decay, $\epsilon$, is
proportional to $U^3/L$, where $U\sim E^{1/2}$ is the typical velocity, we have
\EQ
-{\dd E\over\dd t}\equiv\epsilon
\sim{U^3\over L}\sim{E^{3/2}\over L}\sim E^{5/2},
\EN
and integration over $t$ yields
\EQ
E\sim t^{-2/3}.
\EN
Although this decay law seems compatible with the numerical results
within the range of magnetic Reynolds numbers they considered, its
validity has been challenged on the grounds that $H(t)$ is not
strictly conserved, but that it too must decay.
Christensson et al.\ (2005) used the fact that $H(t)$ obeys the decay law
\EQ
\dot{H}=-2\eta k_{\rm d}^2H,
\EN
where $2\pi/k_{\rm d}\equiv\ell_{\rm d}$ is the typical scale on
which magnetic helicity dissipation occurs.
The decay law of $H(t)$ can only have power law behavior if
$k_{\rm d}$ scales like $k_{\rm d}\sim t^{-1/2}$.
We make use of this assumption and
write this relationship in the following more explicit form:
\EQ
k_{\rm d}=k_{\rm d0}\left(t/t_0\right)^{-1/2},
\EN
where $k_{\rm d0}$ and $t_0$ are suitably defined constants.
With this we have
\EQ
H\sim t^{-2s},\quad\mbox{where}\quad
s=\eta k_{\rm d0}^2t_0.
\EN
Simulations show that a number of different length scales,
including $\ell_{\rm d}$ and $L$, are all proportional to each
other, and that their ratios are independent of time.
Since $E=H/L$, we find that the energy decay law is
\EQ
E\sim t^{-2s-1/2}.
\EN
The correction to the exponent, $2s$, vanishes in the limit of
large magnetic Reynolds numbers, so that for all practical purposes
the energy decay law is $E\sim t^{-1/2}$.
The same scaling law, but without the correction term for finite
magnetic Reynolds numbers, has also been obtained by Campa\-nelli (2004)
using different scaling arguments.
For comparison with simulations, however, the finite magnetic Reynolds
number correction can be important.
Empirically, Christensson et al.\ (2005) found that $s\approx25/R_{\rm m}$.

We summarize this section by stressing once more the particular importance
played by the magnetic helicity equation and, more specifically, the
resistively slow evolution of the magnetic helicity for large magnetic
Reynolds numbers.
Obviously, the magnetic helicity only plays a role if $H$ is indeed finite
and in fact large enough.
The question ``how large is large?" has not yet been addressed, because
most studies assume the field to be maximally helical.
This means that the magnetic helicity spectrum obeys $|kH(k)|=2E_{\rm M}$,
i.e.\ the realizability condition is saturated.
However, even if the fractional magnetic helicity is initially small,
because $E$ tends to decay faster than $H$, the fractional magnetic
helicity will gradually increase (Vachaspati 2001).

A more serious problem is whether significant levels of magnetic
field strengths can be generated.
The general consensus is now that it may be difficult, albeit not
impossible, to have still a field strength of around $10^{-9}\G$
at the present time.
Such a field might have led to measurable polarization in the cosmic
microwave background (Subramanian \& Barrow 1998b, 2002;
Seshadri \& Subramanian 2001; Mack et al.\ 2002; Lewis 2004).
It may also be possible to detect the presence of magnetic helicity
through the production of a parity-odd component of gravity waves,
which induces parity-odd polarization signals (Caprini et al.\ 2004;
Kahniashvili \& Ratra 2005).
A $10^{-9}\G$ field would also provide a sufficiently powerful seed magnetic
field for explaining the generation and maintenance of fields with
equipartition field strength.
This will be discussed in the next section.

\section{Magnetic helicity in dynamos}

Before we focus specifically on the importance of magnetic helicity in
dynamos, we discuss first whether in the dynamo scenario a significant
magnetic field strength can be generated.
The overall problem lies in the fact that the time scale, on which a
global ordered magnetic field on the scale of galaxies can be generated,
is likely to be comparable to the age of galaxies.
To be successful, one has to have a strong enough seed magnetic field
(Rees 1987).
Typical $e$-folding times are on the order of the rotation period,
which is around $2\times10^8\yr$; see Beck et al.\ (1996).
Such times may be too long in view of the fact that in some very young
high redshift galaxies (age $10^9\yr$) typical field strengths are
already in the microgauss range (Kronberg et al.\ 1992;
Perry et al.\ 1993; Kronberg 1994).
Within a time as short as 5 $e$-folding times one would only be able
to amplify the field by a factor of 150.

\begin{figure*}[t!]\begin{center}
\includegraphics[width=.75\textwidth]{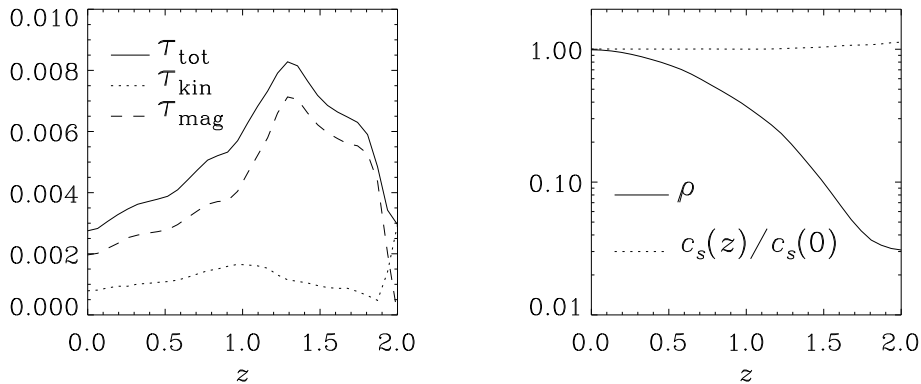}
\end{center}\caption[]{
Dependence of the stress component $\overline{\Pi}_{\varpi\phi}^{\rm(SS)}$
(here denoted by $\tau_{xy}$), separately for the kinetic and magnetic
contributions, together with the sum of the two denoted by total (left)
as well as the vertical dependence of density and sound speed (right).
Note that $\tau_{xy}$ is neither proportional to the density $\rho$ nor
the sound speed $c_{\rm s}$.
[Adapted from Brandenburg et al.\ (1996b)]
}\label{palpz}\end{figure*}

\subsection{Outflows from AGNs or YSOs for seeding galaxies}

In addition to primordial magnetic fields, a potentially much stronger
source of seed magnetic fields might be provided by stellar winds and the
outflows of protostellar discs around young stellar objects (YSOs),
as well as discs around active galactic nuclei (AGNs).
AGNs might provide more coherent fields because their scale is larger.
The general idea of outflows seeding the interstellar medium has
been around for some time (see, e.g., Goldshmidt \& Rephaeli 1993, 1994;
V\"olk \& Atoyan 1999; Brandenburg 2000; Kronberg et al.\ 2001).
On the average the coherence scale of the field in clusters of galaxies
is $5\kpc$ (Clarke et al.\ 2001), but locally it can be much larger.
Along some of the well developed radio lobes the coherence scale
can be as big as a megaparsec (Govoni et al.\ 2001).

In order to estimate the resulting field strength, let us here
reproduce an earlier estimate by Brandenburg (2000).
The basic idea is that outflows (both on stellar and on galactic
scales) tend to be magnetized.
Their power or kinetic luminosity is roughly
\EQ
L_{\rm kin}\approx\dot{M}_{\rm w}c_{\rm s}^2,
\EN
where $\dot{M}_{\rm w}$ is the mass loss rate into the wind
and $c_{\rm s}$ is the sound speed of the ambient gas.
The outflow speed tends to be a certain multiple of this.
Assuming that the ratio of magnetic to kinetic luminosities,
$L_{\rm mag}/L_{\rm kin}$, is
constant [about $0.05$ in the work of von Rekowski et al.\ (2003)]
we can estimate the mean injection of magnetic fields to a cluster
with $N$ sources, each working over a time span $\Delta t$, distributed
over a total volume $L^3$.
This gives a magnetic energy of
\EQ
E_{\rm mag}=N L_{\rm mag} \Delta t
\EN
for the entire cluster, and a root-mean-square field strength of
\EQ
\bra{\BB^2}^{1/2}=\left(8\pi E_{\rm mag}/L^3\right)^{1/2}
\EN
in cgs units.

Assuming $\dot{M}_{\rm w}=0.1M_\odot\yr^{-1}\approx10^{25}\g\s^{-1}$ for
an AGN disc, $c_{\rm s}=1000\km\s^{-1}$ for a galaxy cluster, we have
$L_{\rm kin}\approx10^{41}\erg\s^{-1}$, and hence
$L_{\rm mag}\approx10^{39}...10^{40}\erg\s^{-1}$.
Assuming $\Delta t=0.1\Gyr$ and $N=10^4$ we have
$E_{\rm mag}\approx10^{59}\erg$ for the entire cluster.
Thus, $\bra{\BB^2}^{1/2}\approx0.3\uG$.

For stellar winds and young stellar objects (YSOs)
we obtain a very similar estimate.
Assuming $\dot{M}_{\rm w}=10^{-8}M_\odot\yr^{-1}\approx10^{18}\g\s^{-1}$
for a disc around a young stellar object, $c_{\rm s}=10\km\s^{-1}$
for the warm interstellar medium, we have
$L_{\rm kin}\approx10^{30}\erg\s^{-1}$, and hence
$L_{\rm mag}\approx10^{28}...10^{29}\erg\s^{-1}$.
Assuming $\Delta t=1\Myr$ and $N=10^{11}$ we have
$E_{\rm mag}\approx10^{53}\erg$ for an entire galaxy.
Thus, again, $\bra{\BB^2}^{1/2}\approx1\uG$.

A potential problem with these approaches is that the magnetized
winds may not actually be able to penetrate much (Jafelice \& Opher 1992).
A completely different idea is to produce strong enough seed magnetic
fields in protogalactic turbulence by the small-scale dynamo, whose
time scale is much shorter ($10^7\yr$); see Beck et al.\ (1994)
for more details.
The small-scale dynamo could produce a significant $\overline{\uu\cdot\bb}$
correlation which would contribute to the $\alpha$ effect
(Yoshizawa \& Yokoi 1993; Brandenburg \& Urpin 1998).
It is also possible that a combination of outflows together with small
scale dynamo action might be providing the necessary seed for the large
scale dynamo.

\subsection{Disc corona heating by the MRI}

In order to drive the outflows that may contribute to seeding the
interstellar medium and that remove small-scale magnetic helicity
from the dynamo (see next section), we need to discuss briefly the
physics of disc coronae from where such outflows emerge.

A highly probable source of turbulence in any accretion disc is the
Balbus \& Hawley (1991) or magneto-rotational instability (MRI).
Simulations show that the MRI together with the dynamo instability
can produce a doubly-positive feedback, sustaining both the
turbulence and the magnetic field necessary to drive the turbulence;
see Brandenburg et al.\ (1995), Hawley et al.\ (1996),
Stone et al.\ (1996).
As has been emphasized in a number of papers, the MRI has the
property of liberating most of its energy in the outer parts of
the disc or rather the disc corona, where the density is low and
the heating per unit mass therefore high.
This was originally demonstrated only for nearly isothermal
discs (Brandenburg et al.\ 1996b), see \Fig{palpz}, but this has
now also been shown for radiating discs (Turner 2004).

The mechanism of heating disc coronae described here is essential
in the aforementioned picture of driving magnetized winds from
accretion discs.
It should however be noted that the conical outflows found by
von Rekowski et al.\ (2003) may actually be more general and have
now also been seen in fully three-dimensional simulations
(De Villiers et al.\ 2005).

\subsection{Importance of outflows for dynamos}

Over the past 10--15 years it has become clear that the original mean
field dynamo theory misses something important regarding its saturation
properties.
It started off by numerical calculations of the diffusion of
a mean magnetic field in two dimensions (Cattaneo \& Vainshtein 1991).
These simulations indicated severe quenching of the turbulent magnetic
diffusivity with increasing magnetic Reynolds number.
This prompted similar investigations of the $\alpha$ effect
(Vainshtein \& Cattaneo 1992).
Catastrophic quenching of the $\alpha$ effect was later confirmed using
three-dimensional simulations (Cattaneo \& Hughes 1996).
Calculations involving magnetic helicity arguments were already presented
by Gruzinov \& Diamond (1994) and Bhattacharjee \& Yuan (1995), confirming
again a magnetic Reynolds number-dependent (i.e.\ catastrophic) $\alpha$
quenching.
Another idea was that a sub-equipartition field would lead to
the suppression of Lagrangian chaos (Cattaneo et al.\ 1996).
The general idea was that small-scale magnetic fields grow rapidly
to equipartition field strength, and that at this point the
$\alpha$ effect shuts off (see also Kulsrud \& Anderson 1992).
Clearly, if mean field theory has anything to do with large-scale
dynamos in galaxies or even the much smaller AGN and YSO discs,
then something must be wrong with the idea of premature or
catastrophic quenching (Field 1996).

Only over the past 5 years it became clear that the real culprit
is indeed the magnetic helicity of the small-scale field, as was
already suggested by Gruzinov \& Diamond (1994, 1995, 1996) and
Bhattacharjee \& Yuan (1995), and that this problem might then
be possible to solve by allowing for outflows of small-scale magnetic
helicity through the boundaries (Blackman \& Field 2000a,b;
Kleeorin et al.\ 2000, 2002, 2003).
The detailed quenching behavior seen in simulations (Brandenburg 2001a)
were important in developing a revised mean field theory
(Field \& Blackman 2002; Blackman \& Brandenburg 2002; Subramanian 2002),
which all have in common an explicitly time-dependent equation for a
magnetic contribution to the $\alpha$ effect.
The resulting explicitly time-dependent equation is virtually identical to
the old time-dependent quenching theory of Kleeorin \& Ruzmaikin (1982).

\begin{figure}[t!]\centering
\includegraphics[width=\columnwidth]{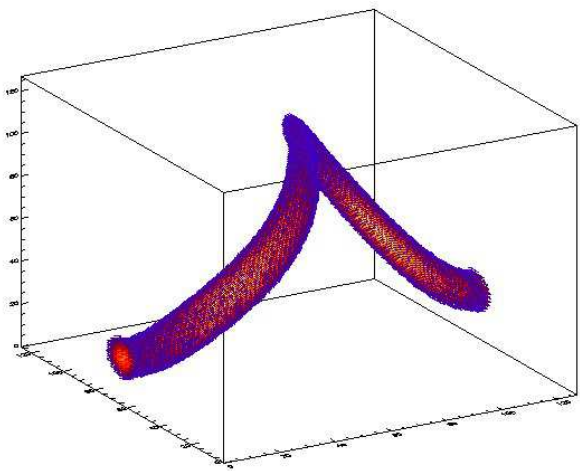}\caption{
Magnetic flux tube constructed from a Cauchy solution
describing analytically the tilting and associated internal
twisting of the tube.
[Adapted from Yousef \& Brandenburg (2003)]
}\label{Fvecgd_t0}\end{figure}

\begin{figure}[t!]\centering
\includegraphics[width=\columnwidth]{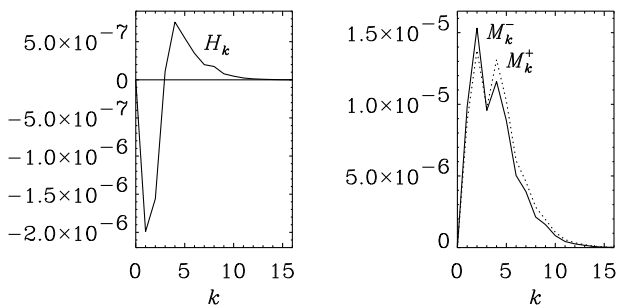}\caption{
Initial spectra of magnetic helicity, $H_k$, and of magnetic energy
of positively and negatively polarized components, $M_k^+$ and $M_k^-$,
respectively, for the tilted and twisted flux tube depicted in \Fig{Fvecgd_t0}.
[Adapted from Yousef \& Brandenburg (2003)]
}\label{Fpower_init}\end{figure}

A pictorial explanation of these new developments can be given as follows.
Stratified rotating turbulence produces cyclonic motions, just as
envisaged by Parker (1955).
This produces in a systematic fashion a tilt in toroidal flux tubes as
they rise owing to either thermal or magnetic buoyancy.
This tilt is the source of producing a poloidal field from a
systematically oriented toroidal field.
However, what was not included in this picture is the fact that
an externally imposed tilt must necessarily yield an internal twist
in the tube (Blackman \& Brandenburg 2003).
This can be seen in a semi-analytically generated
Cauchy solution of an initially straight tube subject to a simple
rising and twisting motion (Yousef \& Brandenburg 2003); see
\Fig{Fvecgd_t0}.
The magnetic helicity spectrum confirms a distinctively bi-helical
behavior; see \Fig{Fpower_init}.
This shows that due to this imposed motion no net magnetic helicity
is produced, and that this is done in such a way that finite magnetic
helicity is being produced with opposite signs at large scales
($H_k<0$) and at smaller scales ($H_k>0$).
The same result was also obtained by Blackman \& Brandenburg (2003),
who calculated numerically the rise, expansion, and subsequent tilt of
a flux tube in the presence of the Coriolis force.

The consequences of producing small-scale magnetic helicity can
be dramatic in some cases (e.g.\ in periodic boxes).
How this works has to do with another development that has its roots
way in the past (Pouquet et al.\ 1976), but whose consequences were
not appreciated until more recently.
The point is that all the analytically derived expressions for the
$\alpha$ effect must be attenuated by an extra term (a magnetic $\alpha$
effect) that is proportional to the small-scale current helicity,
$\overline{\jj\cdot\bb}$, in the isotropic case, or a corresponding
modification proportional to $\epsilon_{ijk}\overline{b_kb_{j,p}}$
in the anisotropic case.
(Here, $\bb=\BB-\meanBB$ is the deviation from the mean magnetic field,
i.e.\ the fluctuating field, and $\jj=\nab\times\bb$ is the fluctuating
current density, where the vacuum permeability is put to unity.)

The reason this term has not been included in the past is that it does
not normally occur in the standard first order smoothing approximation
that has frequently been used for calculating the $\alpha$ effect.
However, when the so-called minimal tau approximation is used
(Blackman \& Field 2002; R\"adler et al.\ 2003;
see review by Brandenburg \& Subramanian 2005a)
this term appears quite naturally.
Kleeorin \& Rogachevskii (1999) have already used the $\tau$ approximation
much earlier, and the $\overline{\jj\cdot\bb}$ correction was also used by
Gruzinov \& Diamond (1994) and Bhattacharjee \& Yuan (1995),
referring to original work of Pouquet et al.\ (1976), who were the first
to use Orszag's (1970) $\tau$ approximation in MHD.
Simulations have now verified explicitly the existence of the
$\overline{\jj\cdot\bb}$ term (Brandenburg \& Subramanian 2005b).

\begin{figure}[t!]\begin{center}
\includegraphics[width=\columnwidth]{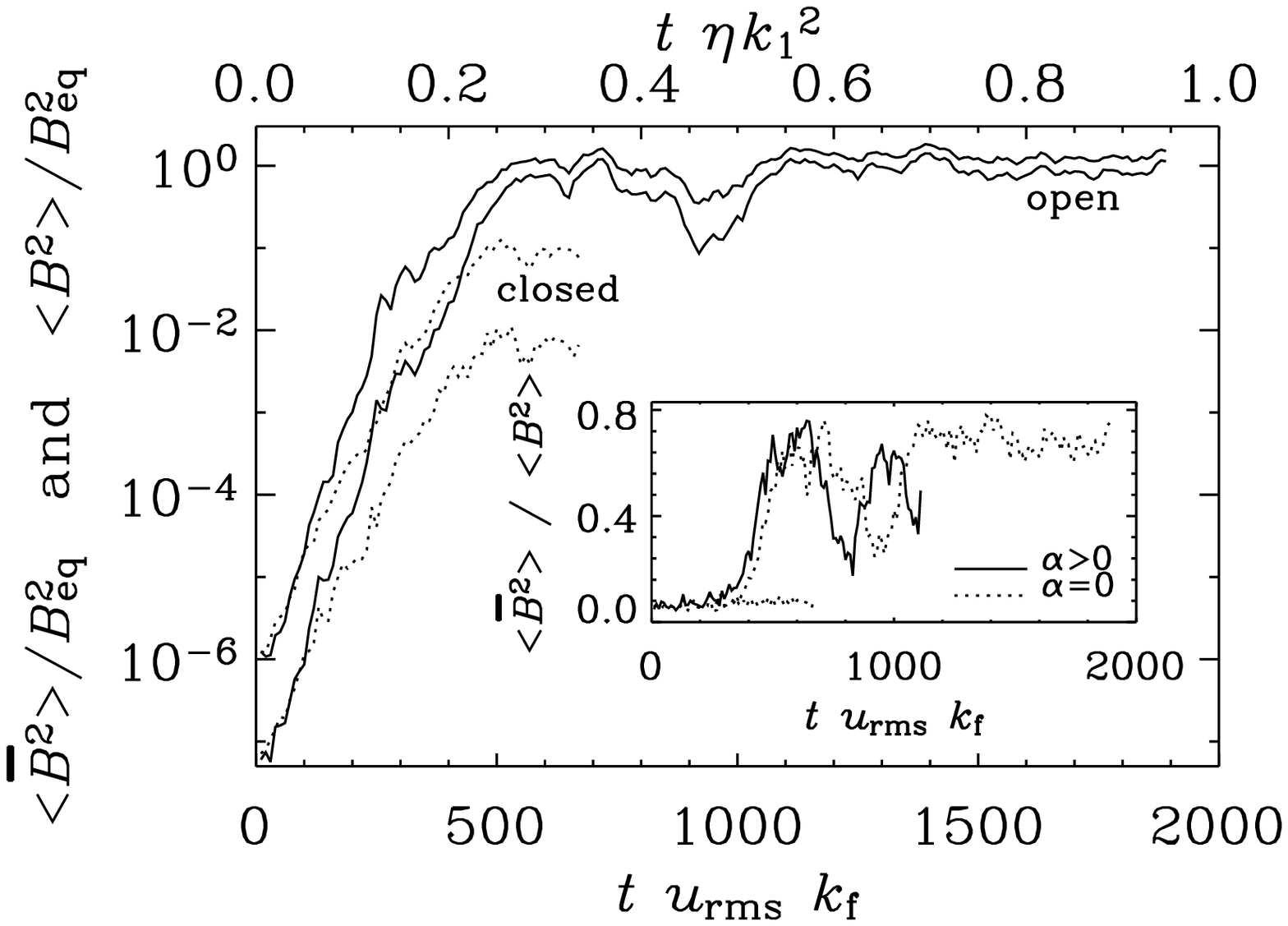}
\end{center}\caption[]{
Evolution of the energies of the total field $\bra{\BB^2}$ and of
the mean field $\bra{\meanBB^2}$, in units of $B_{\rm eq}^2$,
for runs with non-helical forcing
and open or closed boundaries; see the solid and dotted lines, respectively.
The inset shows a comparison of the ratio $\bra{\meanBB^2}/\bra{\BB^2}$
for nonhelical ($\alpha=0$) and helical ($\alpha>0$) runs.
For the nonhelical case the run with closed boundaries is also
shown (dotted line near $\bra{\meanBB^2}/\bra{\BB^2}\approx0.07$).
Note that saturation of the large-scale field occurs on a
dynamical time scale; the resistive time scale is given on the
upper abscissa.
[Adapted from Brandenburg (2005)]
}\label{pmean_comp}\end{figure}

Once the proper course of the catastrophic quenching phenomenon was
discovered, it became relatively easy to identify possible remedies,
such as the allowance for helicity fluxes.
However, it is not enough to allow for open boundaries; e.g.\ in a box
with open boundaries such quenching yields saturation field strengths
that depend on the magnetic Reynolds number (Brandenburg \& Dobler 2001).
It is necessary to have, throughout the domain, an active driver of
helicity flux (magnetic or current helicity), for example shear
(Vishniac \& Cho 2001; Subramanian \& Brandenburg 2004, 2006).
In \Fig{pmean_comp} we demonstrate the importance of open versus
closed boundaries in a simulation of forced turbulence with shear
(Brandenburg 2005).
The simulation shown here has a shear profile that is relevant to a
local model of part of the solar convection zone, but it is expected
that the same physics carries over to large-scale dynamo action in
accretion discs.

\section{Conclusions}

Not all magnetic fields will be helical, but if they are, this can
have dramatic consequences for their evolution.
The effects can be equally dramatic both in decaying and in driven
turbulence, as has been highlighted in this review.
Although we have not discussed this in the present paper, it should
be emphasized that helical large-scale magnetic fields can also be generated
in non-stratified shear flows where there is no $\alpha$ effect, but
there can instead be the so-called shear--current of $\meanWW\times\meanJJ$
effect (Rogachevskii \& Kleeorin 2003, 2004).
This effect may also explain the large-scale dynamo action seen in
\Fig{pmean_comp}, where the results without helicity are quite similar
to those with helicity (Brandenburg 2005).
One-dimensional mean field calculations with the $\meanWW\times\meanJJ$
effect (Brandenburg \& Subramanian 2005c) show that in this case a
magnetic $\alpha$ effect can be produced that has different signs on
the two sides of the midplane.
This magnetic $\alpha$ effect thus contributes to the saturation of the
dynamo even if there is no ordinary (kinetic) $\alpha$ effect.
This highlights once more the dramatic effects played by magnetic helicity.

Whether or not the primordial magnetic field was really helical remains
a big question.
If it was, it is likely that an inverse cascade process has produced
fields of progressively larger scale.
This might lead to observable effects in the cosmic microwave background.
Such a field may also be important for seeding the galactic dynamo, but it
is important to realize that a variety of astrophysical mechanisms may
also produce seed fields just as large.
Our estimate for magnetized outflows from AGNs or YSOs assumes that
the source remains active for a certain period of time, and that their
exhaust goes freely into the ambient medium.
Partial evidence for this actually happening lies in the fact that
clusters of galaxies are chemically enriched with heavier elements.
Given that magnetic fields are intrinsically connected with the outflow,
just like the heavier elements in it, it is quite plausible that
some degree of magnetic contamination of the cluster must have occurred.

In order to produce finally the observed large-scale magnetic fields of
galaxies, some more reshaping, amplification, and maintenance against
magnetic decay is necessary.
Roughly, we expect this to happen just like the mean field dynamo is
able to amplify and maintain the field, although it must operate
on an already strong enough field.
This initial field will still be random and of mixed parity about the
midplane (or equator), but there will be some finite degree of quadrupolar
field which is the one that is dominant in many galaxies;
see Krause \& Beck (1998) and Brandenburg \& Urpin (1998)
for a related discussion about the importance of seeding the quadrupolar
field component.
As we have argued above, the catastrophic quenching problem of the
dynamo has to be overcome, and this is likely to be the case because
of various magnetic and current helicity fluxes operating within the
entire dynamo domain.
In the context of the solar dynamo, simulations have now begun to demonstrate
the dramatic difference made by open boundary conditions, and we hope that
a similar demonstration will soon be possible for the galactic dynamo as well.
Corresponding mean field calculations have already been performed showing
that the catastrophic quenching effect is overcome by an advective flux out
of the domain along the vertical direction.
In particular, it will be interesting to see whether the shedding of magnetic
helicity can actually lead to directly observable effects.

\acknowledgements
I thank Eric Blackman, Mark Hindmarsh, and Kandaswamy Subramanian
for suggestions and comments on the manuscript.
The Danish Center for Scientific Computing is acknowledged for granting
time on the Horseshoe cluster.

\vfill\bigskip\noindent\tiny\begin{verbatim}
$Header: /home/brandenb/CVS/tex/mhd/bologna/paper.tex,v 1.60 2006/01/23 15:33:09 brandenb Exp $
\end{verbatim}

\end{document}